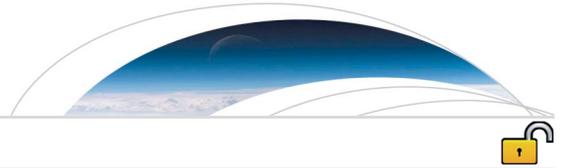

# Mapping lightning in the sky with a mini array

**Martin Füllekrug[1], Zhongjian Liu[1], Kuang Koh[1], Andrew Mezentsev[2], Stéphane Pedeboy[3], Serge Soula[4], Sven-Erik Enno[5], Jacqueline Sugier[5], and Michael J. Rycroft[1]**

[1]Department of Electronic and Electrical Engineering, Centre for Space, Atmospheric and Oceanic Science, University of Bath, Bath, UK, [2]Birkeland Centre for Space Science, University of Bergen, Bergen, Norway, [3]Méteorage, Pau, France, [4]Laboratoire d'Aérologie, Université de Toulouse, Toulouse, France, [5]Remote Sensing and Aircraft Based Observation R and D, MetOffice, Exeter, UK

**Abstract** Mini arrays are commonly used for infrasonic and seismic studies. Here we report for the first time the detection and mapping of distant lightning discharges in the sky with a mini array. The array has a baseline to wavelength ratio $\sim 4.2 \cdot 10^{-2}$ to record very low frequency electromagnetic waves from 2 to 18 kHz. It is found that the mini array detects ~69 lightning pulses per second from cloud-to-ground and in-cloud discharges, even though the parent thunderstorms are ~900–1100 km away and a rigorous selection criterion based on the quality of the wavefront across the array is used. In particular, lightning pulses that exhibit a clockwise phase progression are found at larger elevation angles in the sky as the result of a birefringent subionospheric wave propagation attributed to ordinary and extraordinary waves. These results imply that long range lightning detection networks might benefit from an exploration of the wave propagation conditions with mini arrays.

## 1. Introduction

Lightning location networks use a set of radio receivers to detect and geolocate cloud-to-ground and in-cloud discharge processes associated with thunderclouds [e.g., *Nag et al.*, 2015; *Rakov*, 2013, and references therein]. The geolocation of a lightning discharge is based on the extraction of a significant feature from its electromagnetic waveform that is found above a trigger threshold. The arrival time differences of this feature between pairs of receivers are subsequently used to determine the source location that is found at the best possible intersection of the hyperbolas with constant time, or distance, differences between receiver pairs [e.g., *Rison et al.*, 2016; *Wu et al.*, 2014; *Stock et al.*, 2014; *Betz et al.*, 2004; *Dowden et al.*, 2002; *Cummins et al.*, 1998; *Mazur et al.*, 1997; *Lee*, 1986; *Lewis et al.*, 1960, and references therein]. The separation between the receivers in these lightning location networks is determined by the service area and the frequency, or wavelength, of the recorded electromagnetic waves. Typical baseline to wavelength ratios vary in the range from ~10 to 10,000 (e.g., compare Tables 1 and 2 in *Nag et al.* [2015]). For example, very large baseline to wavelength ratios are associated with global lightning detection networks [*Mallick et al.*, 2014; *Said et al.*, 2013; *Jacobson et al.*, 2006; *Rodger et al.*, 2006]. More recently, lightning location networks with baseline to wavelength ratios ~1–2 were developed [*Lyu et al.*, 2014; *Mezentsev and Füllekrug*, 2013; *Bitzer et al.*, 2013]. One novel application of these lightning detection networks is the remote sensing of high peak current in-cloud lightning discharges and initial stepped leaders that might be associated with the generation of terrestrial gamma ray flashes [*Lyu et al.*, 2016, 2015; *Cummer et al.*, 2015, 2014; *Stanley et al.*, 2006]. In addition, a small aperture array with an extremely small baseline to wavelength ratio $\sim 8.4 \cdot 10^{-2}$ has been used to map radio noise sources in the sky [*Füllekrug et al.*, 2015a, 2014]. This novel technique enabled the identification of a birefringent lower *D* region ionosphere that causes a splitting of the wave propagation path, where the ordinary wave is unaffected by the geomagnetic field and the extraordinary wave penetrates deeper into the ionosphere at larger altitudes as a result of its interaction with the geomagnetic field [*Budden*, 1961]. Experimental evidence for ordinary and extraordinary very low frequency radio waves was recently inferred from the resolution of two distinct source locations in the sky separated by central angles ~0.2°–1.9° [*Füllekrug et al.*, 2015b]. This angular resolution was found to be limited by fluctuations in the lower *D* region ionosphere on the microsecond time scale rather than by the fundamental limit imposed by the timing accuracy of the radio receivers in the array. The aim of this contribution is to identify practical implications of this initial work for mapping lightning discharges in the sky, i.e., to determine whether birefringent subionospheric wave propagation plays a role in long range lightning detection, where the lightning discharges are >1000 km away.

## 2. Mini Array

This study uses a mini array to explore the physical limits of long range lightning detection. The array consists of $N=10$ radio receivers that are randomly scattered across an area ~1 km$^2$ on Charmy Down airfield near Bath in the United Kingdom. The irregular spacing between the instruments results from logistic constraints and accessibility to the area. The largest distance between a receiver pair is 1.26 km, i.e., the baseline $b$ of the array [*Mezentsev and Füllekrug*, 2013]. The mini array measures the electric field strength with a sampling frequency of 1 MHz, i.e., at a temporal resolution of 1 μs with an accuracy ~20 ns [*Füllekrug*, 2010]. Exemplary measurements on 13 May 2011, from 15:00:00 to 15:00:10 UT are extracted from the original recordings. The seemingly short duration of 10 s is chosen because it enables an analysis without the need for high performance computing. The recordings are used to investigate transverse electromagnetic waves that sweep across the array in $\Delta t = b/v = 4.2$ μs when $v$ is the speed of light $c$, but the electromagnetic waves can also appear to the array with an apparent velocity $v = c/\cos\Theta$ when they arrive from an elevation angle $\Theta$ such that $\Delta t = b \cos\Theta/c < 4.2$ μs [*Rost and Thomas*, 2002]. The measurements are filtered with a bandwidth of ±8 kHz around the center frequency $f = 10$ kHz with a wavelength $\lambda \approx 30$ km such that the baseline to wavelength ratio is $b/\lambda \approx 4.2 \cdot 10^{-2}$. This type of mini array with an extremely small aperture enables a sampling of the wavefront with subwavelength precision because one wavelength sweeps across the array in ~100 μs. The phases of the electromagnetic waveforms are referenced to the start time of the recordings by multiplication with $e^{-i\omega t}$, where $\omega = 2\pi f$ is the radian center frequency. The resulting complex electric field strengths $y_n$ inferred from the $n$th receiver at the location $r_n$ is then related to the electromagnetic source field $E$ and the wave model by

$$y_n(t) = E(t)e^{-i\mathbf{k}(t)\mathbf{r}_n}, \quad n = 1, 2 \ldots N. \quad (1)$$

The solution of this system of equations results in the wave number vector $\mathbf{k}(t)$, the electromagnetic source field $E(t) = |E(t)|e^{i\phi(t)}$ in amplitude $|E(t)|$ and phase $\phi(t)$, and the spatial coherency $coh(t)$ of the electromagnetic source field across the array for each individual time step with a resolution of 1 μs

$$coh = \left|\frac{1}{N}\sum_{n=1}^{N}\frac{y_n e^{i\mathbf{k}\mathbf{r}_n}}{|y_n e^{i\mathbf{k}\mathbf{r}_n}|}\right| = \left|\frac{1}{N}\sum_{n=1}^{N}\frac{E_n}{|E_n|}\right| = \left|\frac{1}{N}\sum_{n=1}^{N}e^{i\phi_n}\right| = \left|\frac{1}{N}\sum_{n=1}^{N}\cos\phi_n + i\sin\phi_n\right| \quad (2)$$

[*Füllekrug et al.*, 2014; *Schimmel and Paulssen*, 1997], where $E_n$ and $\phi_n$ are the electromagnetic source field and its phase at each receiver after a correction of the original recordings for the phase delay $\mathbf{k}(t)\mathbf{r}_n$ caused by the time varying wave number vector. A spatial coherency of 1 means that the source field is a plane wave that is consistently recorded at each individual receiver across the entire array. Lower spatial coherencies indicate small scale ripples along the wavefront that are caused by either the source mechanism or the varying properties of the medium along the wave propagation path from the source to the array.

## 3. Lightning Detection

Transient electromagnetic pulses from distant lightning discharges occur intermittently, superimposed on a more continuous electromagnetic background radiation from a variety of origins. Thus, an event is detected by a trigger that discriminates the wanted lightning signal against the unwanted background. For example, a trigger could use an amplitude threshold or the spatial coherency of the electromagnetic source field.

### 3.1. Amplitude Threshold Trigger

The envelope of the electromagnetic source field $|E| = \sqrt{E_{re}^2 + E_{im}^2}$ is composed of the real and imaginary part $E_{re}$ and $E_{im}$. These envelope amplitudes exhibit relative maxima caused by lightning discharges (Figure 1, left). The significance of the relative maxima can be estimated by comparison with an amplitude distribution calculated from a purely random process. Here we simulate a random amplitude distribution of the electromagnetic source field by use of the source field spectrum at each receiver. While the spectral amplitudes of $E_n$ are preserved, the phases of the spectrum are scrambled by a random permutation prior to an inverse Fourier transform to calculate the simulated source field $\hat{E}_n$. The resulting random amplitudes are Rayleigh distributed

$$f(|\hat{E}|, \sigma) = \frac{|\hat{E}|}{\sigma^2} \exp\left(-\frac{|\hat{E}|^2}{2\sigma^2}\right), \tag{3}$$

where $\sigma$ is the standard deviation of $\hat{E}_{re}$ and $\hat{E}_{im}$ because both are normally distributed. The comparison with the observed amplitude distribution $f(|E|)$ shows that all amplitudes >85 μV/m from lightning pulses are more likely to occur than the amplitudes >85 μV/m from a random amplitude distribution. This cross over point could be used as an amplitude threshold trigger [*Chrissan and Fraser-Smith*, 2000]. In this case, 35% (~0.9 $\sigma$) of the observed amplitudes would be significant. A more stringent amplitude threshold would be to use the largest 10% (~1.6 $\sigma$) of the amplitude distribution which corresponds to an amplitude threshold >250 μV/m (Figure 1, left). However, neither of these amplitude thresholds takes into account the quality of the signal that is determined by the spatial coherency of the source field across the array.

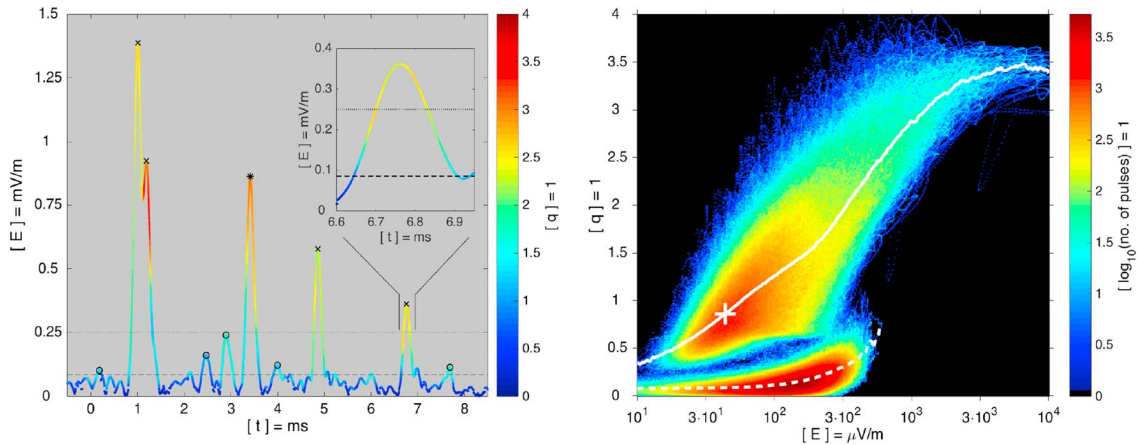

**Figure 1.** Electromagnetic source field and its quality. (**left**) Transient electromagnetic pulses from lightning discharges cause electric field strengths with an envelope that exhibits distinct relative maxima with a varying degree of prominence. Ten relative maxima exceed a threshold of 85 μV/m and five maxima exceed 250 μV/m. Larger relative maxima generally exhibit larger wavefront qualities (line color). Five relative maxima have a quality between 1 and 2 (circles), (four relative maxima have a quality between 2 and 3 (crosses), and one relative maximum has a quality between 3 and 4 (star). It is surprising that the pulse at ~3.4 ms has a larger quality than the pulse at ~1.0 ms even though it has a smaller amplitude. The maximum of the secondary pulse at ~ 1.2 ms has a quality of ~2.95 slightly below the quality threshold while some parts of the rising and falling edge of the waveform have qualities >3. This is possible because the quality can vary during individual lightning pulses (inset). (**right**) The wavefront qualities of lightning pulses vary in the range from ~0.5 to 4 and the average quality generally increases with the electric field strength (solid line). The most likely quality is found near the fix point for the quality of the spatial coherency (plus). The average quality from a simulated random process is calculated for comparison and it varies only slightly with the electric field strength (dashed line).

### 3.2. Spatial Coherency Trigger

It is found that the transient electromagnetic pulses from lightning discharges recorded with the mini array typically have spatial coherencies that vary in the range from 0.9000 to 0.9999 because the wavefronts exhibit little variation across the mini array. These spatial coherencies are much larger than the spatial coherencies that are expected from random spatial coherencies as can be inferred from the following assessment. The spatial coherencies of a random process are determined by uniformly distributed phases $\phi_n$ in equation (2) that result in normally distributed real and imaginary parts $\cos\phi_n$ and $\sin\phi_n$ with a standard deviation $1/\sqrt{2}$. This standard deviation is further reduced by $1/\sqrt{N}$ when the real and imaginary parts are averaged over $N$ receivers such that the final standard deviation is $\sigma_{coh} = 1/\sqrt{2N} \approx 0.22$ for $N = 10$. The absolute value of the coherency calculated from the averaged real and imaginary parts is then Rayleigh distributed with the most likely coherency $coh_{max} = \sigma_{coh}$ and the mean coherency $coh_{mean} = \sigma_{coh}\sqrt{\pi/2} \approx 0.28$, both of which are much smaller than the observed spatial coherencies for the transient electromagnetic pulses from lightning discharges. It is therefore convenient to define a measure for the quality $q$ of the spatial coherency of the electromagnetic source field across the array, i.e., how flat the wavefront is

$$q = \log_{10}\frac{1}{1-coh} = -\log_{10}(1-coh). \tag{4}$$

This quality of the wavefront translates the measured spatial coherencies from 0.9000 to 0.9999 to quality values from 1 to 4 (Figure 1). It is interesting to note that equation (4) is transcendental with a fix point for the quality $q_c$ = coh that has the solution $q_c \approx 0.86$ obtained from a fixed point iteration (Figure 1, right). The quality of this fix point is used as a reference trigger threshold for the detection of lightning pulses with the mini array.

### 3.3. Comparison of Amplitude and Quality

The amplitude threshold trigger can be compared to the quality trigger by use of their bivariate distribution function (Figure 1, right). It is found that the quality generally increases with increasing amplitudes. The maximum of this bivariate distribution function is coincidentally found close to the fix point $q_c$ of the transcendental equation (4). From this comparison alone, the use of either threshold trigger might be considered to be sufficient for lightning detection. However, it is more rigorous to compare the bivariate distribution function to the bivariate distribution function inferred from the simulated randomized electromagnetic source field calculated in section 3.1. In particular, this bivariate distribution function helps to estimate the significance $s$ of possible trigger thresholds by use of the inverse error function

$$s = \sqrt{2}\,\mathrm{erf}^{-1}\left(1 - \frac{n}{N}\right), \tag{5}$$

where erf is the error function [*Arfken and Weber*, 2005], $n$ is the number of simulated random samples that exceed the trigger threshold, $N$ is the total number of random samples, and $p = n/N$ is the probability to exceed the trigger threshold. For example, the fix point $q_c$ corresponds to a significance at the ~3.8 $\sigma$ level, or the chance of one out of 6911 qualities from the simulated randomized electromagnetic source field to exceed $q_c$. The same significance ~3.8 $\sigma$ level is obtained for an amplitude trigger threshold ~0.5 mV/m that would unnecessarily disqualify a large number of signals with high quality wavefronts. As a result, it is more rigorous to base the detection of lightning pulses solely on the quality of the wavefronts, i.e., without using amplitude trigger thresholds, to map lightning discharges in the sky as described in the following section.

## 4. Lightning Mapping

Transient electromagnetic pulses from lightning discharges are extracted from the recordings when the relative amplitude maxima exhibit large wavefront qualities that vary in the range from 3 to 4 such that their significance is well beyond any doubt. The prominence of concurrent relative amplitude maxima is used to reject a very minor fraction (2.8%) of lightning pulses from the subsequent analysis. During the recording of 10 s length, 692 transient electromagnetic pulses from lightning discharges are detected. The pulse rate of ~69 per second is larger than the global lightning flash rate ~44 per second [*Christian et al.*, 2003] because a lightning flash could be composed of multiple strokes. In addition to the relative maximum, concatenating samples that precede the relative maximum on the rising edge and samples that follow the relative maximum on the falling edge are also used for the mapping of lightning pulses in the sky if their qualities vary in the range from 3 to 4. On average, ~150 concatenating samples per maximum are detected such that ~103.8 thousand samples are included for the mapping of lightning pulses in the sky by use of their wave numbers.

### 4.1. Mapping in the Wave Number Domain

The wave numbers inferred from equation (1) are measured in the east-west and north-south directions $k_e$ and $k_n$ such that $k = \sqrt{k_e^2 + k_n^2}$. These wave numbers are subsequently normalized $\kappa_e = k_e/k_m$ and $\kappa_n = k_n/k_m$ with respect to the limiting wave number $k_m = k_0 \cdot 1.25$, where $k_0 = \omega/c$ is the reference wave number for electromagnetic waves in vacuum, $\omega$ is the radian center frequency of the wave, $c$ is the speed of light, and the constant 1.25 is inferred from the lowest elevation angles observed in the sky [*Füllekrug et al.*, 2015a]. The wave numbers are finally mapped on individual pixels in the normalized wave number plane $\kappa_e \times \kappa_n$ with a resolution $\Delta\kappa = 0.1$. This cumulative two-dimensional distribution function of lightning pulses $F$ in the wave number plane exhibits a scatter of natural origin that is much larger than the small scatter imposed by the timing accuracy of the radio receivers in the array [*Füllekrug et al.*, 2015b]. The observed scatter can be reduced by use of an impulse response, or point spread, function $H$ that is determined by the standard deviation $\sigma_\kappa$ of the normalized wave numbers

$$H(\boldsymbol{\kappa}, \sigma_\kappa) = \frac{1}{2\pi\sqrt{|\mathbf{C}|}} \exp\left(-\frac{1}{2}\boldsymbol{\kappa}^T \mathbf{C}^{-1} \boldsymbol{\kappa}\right), \tag{6}$$

where $\mathbf{C}$ is the covariance matrix with $\sigma_\kappa^2$ as diagonal elements and $|\mathbf{C}|$ is its determinant. The convolution of $F$ with the inverse of the point spread function $H^{-1}$ results in a sharpened image $Z$ in the wave number domain $Z = F * H^{-1}$ that is more conveniently calculated as a fraction $z = f./h$, where $./$ is the element-wise matrix division and the lower case symbols represent quantities in the two-dimensional Fourier domain. For example, the image $Z$ is represented in the Fourier domain by

$$z(u,v) = \sum_{n=0}^{N-1} \sum_{m=0}^{M-1} Z(n\Delta\kappa, m\Delta\kappa) e^{-i2\pi(un/N + vm/M)} \quad (7)$$

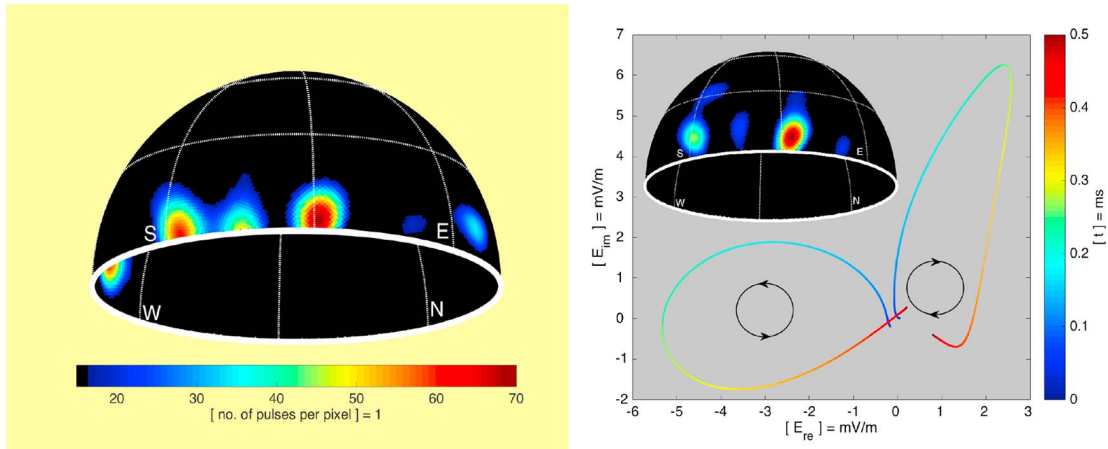

**Figure 2.** Maps of lightning discharges in the sky. **(left)** The mapped lightning pulses occur at several arrival azimuths that are associated with prominent thunderstorms in northern Spain, southern France and Italy and smaller storms over the Atlantic and the North Sea. Some lightning pulses occur at elevation angles above the horizon (solid line). The longitudes are separated by 45° (vertical dotted lines) with geographic directions labeled 'E', 'N', 'W' and 'S'. Two longitudes cannot be seen because they fall on the left and right edge of the hemisphere as a result of the perspective. The latitudes are located at the elevation angles 60° and 82° (horizontal dotted lines) as inferred from the coordinate transformation used for the mapping. **(right)** Most lightning pulses exhibit a counterclockwise phase progression in the constellation diagram of the real and imaginary part of the electromagnetic source field, but 14% of the lightning pulses exhibit a clockwise phase progression (colored lines). Lightning pulses with a clockwise phase progression occur at larger elevation angles (inset) when compared to all lightning pulses (compare to left). The larger elevation angles are attributed to extraordinary electromagnetic waves that are affected by the geomagnetic field.

[*Gonzalez and Woods*, 2008].

### 4.2. Mapping in the Sky

The arrival azimuths of the electromagnetic waves emitted by lightning discharges are determined by $\Phi = \arctan(k_n/k_e) = \arctan(\kappa_n/\kappa_e)$ measured in the interval $[-\pi, +\pi]$ with respect to the east direction. In theory, the elevation angle $\Theta$ of electromagnetic waves emitted by lightning discharges is determined by $\cos\Theta = k/k_m = \kappa$ [*Füllekrug et al.*, 2015a; *Rost and Thomas*, 2002]. In practice, however, the wave numbers are determined with an unconstrained Gaussian method of least squares such that $\kappa$ is normally distributed with numerical values $\kappa > 1$. As a result, $\cos\Theta$ is normally distributed such that the elevation angles inferred from the nonlinear transformation $\Theta = \arccos(\kappa)$ are not normally distributed and are imaginary for $\kappa > 1$. It is therefore advantageous to choose a transformation of coordinates to map lightning discharges in the sky. For example, the transformation $\kappa' = 1 - \kappa$ maps the range $\kappa \in [0,2]$ to the interval $\kappa' \in [1,-1]$ such that $\Theta' = \pi/2 - \arccos\kappa'$ is an almost linear measure for 88% of the interval with the remaining 12% at large elevation angles > 70° where no signatures from lightning discharges occur during the observation period. The resulting map of lightning in the sky exhibits five centers of lightning activity that correspond to prominent thunderstorms in northern Spain, southern France, and northern Italy and smaller storms over the Atlantic near Ireland and the North Sea near Denmark (Figure 2, left), as confirmed by lightning detection networks operated by Météorage in France and the MetOffice in the UK around the time of the recordings [*Mezentsev and Füllekrug*, 2013].

### 4.3. Wave Propagation

The mapping of lightning discharges with the mini array offers the unique opportunity to study the medium in which the electromagnetic waves from lightning discharges propagate. For example, some lightning pulses exhibit a clockwise phase progression in the constellation diagram, whereas other lightning pulses exhibit a counterclockwise phase progression (Figure 2, right). The direction of the phase progression is determined by

d$\phi$/dt = $\omega_a$ = 2$\pi f_a$, where $f_a$ is the apparent, or instantaneous, frequency measured relative to the center frequency 10 kHz. For example, an electromagnetic wave with a dominant frequency of 12 kHz has an apparent frequency of 2 kHz with a counterclockwise phase progression and an electromagnetic wave with a dominant frequency of 8 kHz has an apparent frequency of −2 kHz with a clockwise phase progression. It is found here that 97 of the 692 recorded lightning pulses with a clockwise rotation (14%), i.e., smaller dominant frequencies, appear at larger elevation angles in the sky when compared to all lightning discharges (Figure 2, right, inset). These larger elevation angles were previously attributed to a birefringent wave propagation where the extraordinary mode is affected by the geomagnetic field such that the electromagnetic waves are reflected at larger altitudes and more strongly attenuated than the ordinary mode that is not affected by the geomagnetic field [*Füllekrug et al.*, 2015b; *Budden*, 1961]. Many lightning pulses exhibit a counterclockwise rotation during the rising edge of the waveform and a clockwise rotation during the falling edge of the waveform. This result is in agreement with the observation that waveforms of lightning discharges often exhibit a leading high frequency component followed by a low frequency component of the waveform [e.g., *Dowden et al.*, 2002, Figure 12]. The results therefore indicate that the trailing sky wave contributions embedded in the electromagnetic waveforms are affected by the geomagnetic field during their reflection at the lower D region ionosphere.

## 5. Summary and Conclusions

A mini array was successfully used for the first time to map distant lightning discharges in the sky and to discriminate between two different wave propagation paths that result from a birefringent lower D region ionosphere. The baseline to wavelength ratio of the mini array ∼4.2 · 10$^{-2}$ is much smaller than what is normally used by lightning detection networks. This unique capability enables a sampling of the wavefront with subwavelength precision to determine the quality of the wavefront which is a rigorous measure of the spatial coherency for electromagnetic source fields sweeping across the mini array. The relatively large number of event detections ∼69/s shows that lightning pulses associated with cloud-to-ground and in-cloud discharges are observed at distances ∼1000 km as previously reported in the scientific literature [e.g., *Enno et al.*, 2016, and references therein]. The results obtained here therefore imply that long range lightning detection networks are affected by wave propagation conditions that can be determined, and perhaps mitigated, by the use of mini arrays toward an improvement of long range lightning detection. Such a long range lightning detection network composed of mini arrays might need to take into account possible distance dependencies of the wavefront quality and apparent frequency that remain to be determined in future studies.

**Acknowledgments**
The work of M.F. is sponsored by the Natural Environment Research Council (NERC) under grants NE/L012669/1 and NE/H024921/1. The work of Z.L. is sponsored by the China Scholarship Council (CSC) contract 201408060073, the UK MetOffice under grant EA-EE1077, and by the University of Bath. The work of K.K. is sponsored by the Engineering and Physical Sciences Research Council (EPSRC) under DTA contract EB-EE1151. M.F. developed the concepts for the data analysis, conducted the measurements together with A.M., and wrote the publication. Z.L., K.K., and A.M. advised on the signal processing. S.P. and S.S. communicated Météorage lightning location data for comparison. S.E. and J.S. communicated UK MetOffice ATD lightning location data for comparison. M.R. advised on the interpretation of the results. M.F. acknowledges helpful discussions with Michael Stock, Stéphane Gaffet, Robert Watson, Ivan Astin, and Adrian Evans. The data used for this publication are available from http://dx.doi.org/10.15125BATH-00243. This work was inspired by the Royal Meteorological Society meeting on Advances in Lightning Detection in Reading, 9 March 2016, the TEA-IS network of the European Science Foundation and the SAINT project of the European Commission (H2020-MSCA-ITN-2016, 722337). The authors thank the reviewers for their assistance to improve the quality of the manuscript.